\documentclass[conference]{IEEEtran}
\IEEEoverridecommandlockouts
\usepackage{cite}
\usepackage{amsmath,amssymb,amsfonts}
\usepackage{algorithmic}
\usepackage{graphicx}
\usepackage{subcaption}
\usepackage{float}
\usepackage{textcomp}
\usepackage{array, makecell}
\usepackage{multicol}
\usepackage{multirow}
\usepackage{xcolor}
\usepackage{pgfplots}
\usepackage{tikz}
\usepackage{booktabs}
\usepackage{tabularx}
\usepackage{placeins}
\usepackage{amssymb}

\newenvironment{customlegend}[1][]{%
        \begingroup
        \csname pgfplots@init@cleared@structures\endcsname
        \pgfplotsset{#1}%
    }{%
        \csname pgfplots@createlegend\endcsname
        \endgroup
    }%
    \def\addlegendimage{\csname pgfplots@addlegendimage\endcsname}

\def\BibTeX{{\rm B\kern-.05em{\sc i\kern-.025em b}\kern-.08em
    T\kern-.1667em\lower.7ex\hbox{E}\kern-.125emX}}
\begin{document}

\title{PCOV-KWS: Multi-task Learning for Personalized Customizable Open Vocabulary Keyword Spotting\\
\thanks{$^*$Corresponding author.}
}

\author{\IEEEauthorblockN{Jianan Pan}
\IEEEauthorblockA{
\textit{Zhejiang University}\\
Hangzhou, China \\
panjian\_an@zju.edu.cn}
\and
\IEEEauthorblockN{Kejie Huang$^*$}
\IEEEauthorblockA{
\textit{Zhejiang University}\\
Hangzhou, China \\
huangkejie@zju.edu.cn}
}
\maketitle

\begin{abstract}
As advancements in technologies like Internet of Things (IoT), Automatic Speech Recognition (ASR), Speaker Verification (SV), and Text-to-Speech (TTS) lead to increased usage of intelligent voice assistants, the demand for privacy and personalization has escalated. In this paper, we introduce a multi-task learning framework for personalized, customizable open-vocabulary Keyword Spotting (PCOV-KWS). This framework employs a lightweight network to simultaneously perform Keyword Spotting (KWS) and SV to address personalized KWS requirements. We have integrated a training criterion distinct from softmax-based loss, transforming multi-class classification into multiple binary classifications, which eliminates inter-category competition, while an optimization strategy for multi-task loss weighting is employed during training. We evaluated our PCOV-KWS system in multiple datasets, demonstrating that it outperforms the baselines in evaluation results, while also requiring fewer parameters and lower computational resources.
\end{abstract}

\begin{IEEEkeywords}
open-vocabulary keyword spotting, speaker verification, personalization, multi-task learning
\end{IEEEkeywords}

\section{Introduction}
Keyword Spotting (KWS) is a pivotal component of modern speech recognition technology, focusing on the identification of specific words or phrases within continuous audio streams. Unlike traditional speech recognition systems that transcribe entire conversations, KWS systems are designed to detect predefined keywords efficiently and with minimal computational overhead. These systems play a crucial role in various applications, from voice-activated assistants like Amazon's Alexa and Apple's Siri to surveillance and emergency response systems. 

Conventional KWS (C-KWS) have primarily focused on recognizing preset keywords that are not tailored to individual users. Open-vocabulary KWS (OV-KWS) is a key to address this challenge, enabling a model to detect arbitrary keywords without prior training on those specific keywords. In custom keyword detection, users can enroll keywords through either audio or text input. Several approaches have been investigated for this purpose. Some OV-KWS models utilize a two-stage method, beginning with acoustic modeling and followed by a complex keyword search phase, the Weighted Finite State Transducer (WFST) has become the predominant method for graph search in these applications~\cite{grarphsearch1, grarphsearch2, grarphsearch3, grarphsearch4}. A common issue with two-stage methods is that the search process requires significant computational resources. 

Now most OV-KWS models are based on end-to-end approach, a classic implementation is Query-by-Example (QbyE), which compares an input speech with an enrolled utterance~\cite{QbyE1, QbyE2, QbyE3, QbyE4, QbyE5, QbyE6}. There are also cross-modal methods that combine text in different ways~\cite{QbyE-T1, QbyE-T2-libri, QbyE-T3-apple, QbyE-T4-phon}, and their advantage is its simplicity for users and its independence from any specific acoustic conditions during the registration process, but this method also has drawbacks: it struggles to handle words with similar sounds and is prone to being mistakenly activated by confusing or phonetically similar phrases. In recent years, some methods based on metric learning~\cite{metric1, metric2, metric3, metric4, metric5, metric6}, which use fixed-dimensional vectors that represent words of varying lengths, have been reported to directly correlate similarity with relative distances within the embedding space. However, these approaches only adapt to open-set keywords, not explicitly considering the user identity. In response to this challenge, certain approaches~\cite{MTL1, MTL2, PK-MTL} have combined KWS and SV through multi-task learning networks, aiming to identify target users alongside keyword detection. Nonetheless, these methods still fall short of offering individual users a personalized experience with freely customizable keywords.

To address this, we propose the PCOV-KWS system, a multi-task learning framework that integrates OV-KWS and SV, to perform personalized KWS, PCOV-KWS not only enables keyword customization akin to the OV-KWS but also excels in discerning the target user's voice from others.


\begin{figure*}[hbtp]
    \centering
    \includegraphics[page=1, width=\textwidth]{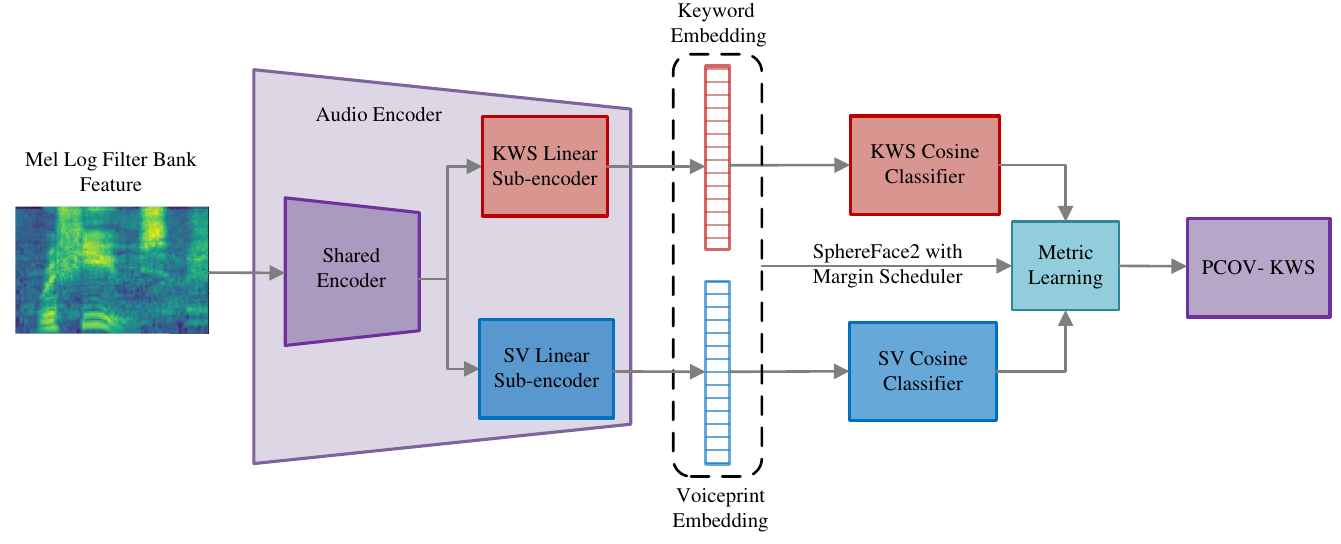}
    \caption{\textbf{Proposed architecture of PCOV-KWS:} The architecture comprises an audio encoder, which includes a shared encoder and two linear sub-encoders for KWS and SV respectively and cosine classifiers integrated with SphereFace 2 for metric learning.}
    \label{fig:mtl_arch}
\end{figure*}

\section{Proposed Approach}
This section outlines the proposed multi-task learning framework for personalized user-defined keyword detection, including the SphereFace2-based~\cite{SphereFace2} metric learning criterion used for training, the loss weighting strategy based on Project Conflicting Gradients (PCGrad)~\cite{PCGrad}, and the large-scale training dataset that we constructed.

\subsection{Multi-task Learning Architecture}
As depicted in Fig.~\ref{fig:mtl_arch}, we propose a multi-task learning architecture that integrates two distinct but partially complementary feature information, KWS and SV, to perform OV-KWS and PCOV-KWS tasks. For multi-task training, the input data are set to multi-label samples, denoted as $\{x_i, y_i^{k}, y_i^{s}\}$, where the $x_i$ represents the input audio feature, and $\{y_i^{k}, y_i^{s}\}$ are the keyword and speaker labels corresponding to each task. 

The audio encoder utilizes hard parameter sharing (HPS)~\cite{HPS} in the bottom layers, which learns the low-level audio features that are common across various tasks. Since the characteristics of KWS and SV are different in the top layers, we separate and copy the top of the encoder to obtain two linear sub-encoders with the same structure to learn the high-level features of each task. Subsequently, the keyword embedding and voiceprint embedding are obtained from the two linear sub-encoders, denoted as $\mathbf{e_i^{k}}$, $\mathbf{e_i^{v}}$. Then there is the cosine classifier based on SphereFace2, which converts the multi-class classification problem into a binary one. This approach is particularly suitable for the OV-KWS task. Given $\mathbf{K}$ classes in the training set, SphereFace2 constructs $\mathbf{K}$ binary classification objectives, treating data from the target class as positive samples and data from all other classes as negative samples. In the case of KWS, cosine classifier $C(\cdot)^{k}$ is shown below:
\begin{equation}
    C(\mathbf{e_i^{k}})^{k} = sim(\mathbf{e_i^{k}}, \mathbf{W_i^{k})},
\end{equation}
where $sim(\mathbf{e_i^{k}}, \mathbf{W_i^{k})}$ is a dot product of normalised $\mathbf{e_i^{k}}$ and the trainable weights of the \textit{i}-th binary classifier $\mathbf{W_i^{k}}$. 

During training, the cosine similarity distribution between positive and negative sample pairs is inconsistent: Negative pairs have a smaller, more concentrated variance, whereas positive pairs exhibit greater variability. This overlap in similarity scores complicates the setting of a clear threshold for distinction. To resolve this, a function is proposed in\cite{SphereFace2} to extend the dynamic range of similarity, computed as:
\begin{equation}\label{eq:gzt}
    g(z) = 2 \left( \frac{z + 1}{2} \right)^t - 1,
\end{equation}
where $z$ is the input consine value and $t$ is a hyperparameter which controls the strength of distribution adjustment.

For a mini-batch of data with $\mathbf{N}$ samples, $\text{where }y_i\in \{1, 2, \ldots, K\}$, the loss is defined as follows:
\begin{equation}
    \mathcal{L}_{k_i}^+ = \lambda \cdot \log \left( 1 + e^{- s \cdot g(C(\mathbf{e_{y_i}^{k}})^{k}, t) + b} \right),
\end{equation}
\begin{equation}
    \mathcal{L}_{k_i}^- = (1 - \lambda) \cdot \sum_{j \neq y_i} \log \left( 1 + e^{s \cdot g(C(\mathbf{e_j^{k}})^{k}, t) + b} \right),
\end{equation}
\begin{equation}
    \mathcal{L}_k = \frac{1}{N} \sum_{i=1}^{N} \left( \mathcal{L}_{k_i}^+ + \mathcal{L}_{k_i}^- \right),
\end{equation}
where s and b indicate scaling factor and bias, and similarly we get $\mathcal{L}_v$.

During the training of the multi-task learning, we are faced with the problem of interference caused by the loss gradient direction conflict between different tasks; here we employ PCGrad for the loss weighting strategy on KWS and SV, which we will compare with Equal Weighting(EW), {\it i.e.}, $\mathcal{L} = \mathcal{L}_k + \mathcal{L}_v$, in the next section. Here we note the gradient of $\mathcal{L}_k$, $\mathcal{L}_v$ as $\mathbf{g_k}, \mathbf{g_v}$, then calculate their inner product:
\begin{equation}
    g_{kv} = \mathbf{g}_k^T \mathbf{g}_v,
\end{equation}
if $g_{kv} < 0$, {\it i.e.}, there is a gradient conflict, correcting it by the following steps:
\begin{equation}
    \mathbf{g}_k \leftarrow \mathbf{g}_k - \frac{g_{kv}}{\|\mathbf{g}_v\|^2 + \epsilon} \mathbf{g}_v,
\end{equation}
where $\epsilon$ is a small value to prevent zero errors. Then we update the loss weight $\omega = [\omega_k, \omega_v]$ (originally [1, 1]):
\begin{equation}
    \omega_k \leftarrow \omega_k - \frac{g_{kv}}{\|\mathbf{g}_v\|^2 + \epsilon},
\end{equation}
similarly, we get the $\omega_v$, then the loss becomes:
\begin{equation}
    \mathcal{L} = \omega_k \cdot \mathcal{L}_k + \omega_v \cdot \mathcal{L}_v
\end{equation}

Inspired by~\cite{PK-MTL}, we incorporate the confidence integration block (CIB) to adapt the MTL model to various tasks, which integrates the confidence in the model output, $\Phi^{k}$ and $\Phi^{v}$, to obtain a new confidence adapted to different tasks during inference. CIB is a defined below:
\begin{equation}\label{eq:CIB}
    \Phi = \alpha \cdot \Phi^{k} + (1-\alpha) \cdot \Phi^{v}.
\end{equation}

\subsection{Large-scale Training Dataset}
Multilingual Spoken Words Corpus (MSWC) is a multilingual keyword dataset that includes more than 23.4 million one-second audio clips corresponding to approximately 340,000 keywords, contributed by roughly 115,000 speakers in 50 languages. In this study, PCOV-KWS model are trained on the English subset of MSWC. Using G2PE, we refine the training data by selecting keywords with more than five phonemes and ensuring a minimum of 30 samples per keyword per speaker. The resulting dataset, after filtering, includes over 1.3 million one-second audio samples, which include 7,757 keywords from 7,908 speakers. 

\subsection{Audio Encoder}

AS shown in Table~\ref{tab:tdres}, the audio encoder is derived from TC-ResNet~\cite{tcresnet}. We refer to the search result obtained~\cite{Autokws} by applying the Noisy Differentiable Architecture Search (NoisyDARTS)~\cite{NoisyDARTS} to TC-ResNet, as well as certain optimization techniques of ConvNeXt V1 and V2, as detailed in~\cite{convnextv1, convnextv2}. Through experimentation, we developed TDResNeXt as our audio encoder, which demonstrates superior network performance and inference efficiency compared to TC-ResNet.

\section{Experiments}

\subsection{Experimental Setups}
\subsubsection{Evaluation Datasets}
We evaluate our PCOV-KWS system with Google Speech Commands v1 ($\textbf{\text{G}}$)~\cite{gsc}, LibriPhrase-easy ($\textbf{\text{LP}}_\textbf{\text{E}}$) and LibriPhrase-hard ($\textbf{\text{LP}}_\textbf{\text{H}}$)~\cite{QbyE-T2-libri} in different scenarios: first, the C-KWS task, in this context, is limited to detecting specific keywords; and second, the OV-KWS task, {\it i.e.}, users can customize any keywords, unlike the C-KWS which is a closed set task; last one, the PCOV-KWS task, designed to recognize keywords that are unique to individual users.
 

\subsubsection{Evaluation metric}
We employ the Equal Error Rate (EER), where FAR equals FRR, and the Area Under the Curve (AUC) as critical metrics for evaluating the performance of KWS models. 


\begin{table}[t]
\caption{Detailed Results for Modifying TC-ResNet} 
\label{tab:tdres}
\begin{center}
\fontsize{9pt}{10pt}\selectfont
\begin{tabular}{|c|c|c|}
\hline
\textbf{Model Modification} & \textbf{MSWC Acc.(\%)} & \textbf{\#FLOPs} \\
\hline
TC-ResNet14-1.5                            & $78.21(0.04)^{\mathrm{a}}$ & 6.86M       \\
stage ratio\{1,1,3,1\}                    & $78.25(0.09)$ & 5.54M       \\
NoisyDARTS                                & $79.48(0.12)$ & 4.63M       \\
"patchify" stem                           & $79.36(0.07)$ & 4.65M       \\
temporal dsconv                           & $78.11(0.07)$ & 1.61M       \\
inverting dimensions                      & $81.45(0.05)$ & 7.87M       \\
move up dsconv                            & $82.18(0.09)$ & 0.92M       \\
kernel size $\Rightarrow$ 5               & $81.91(0.14)$ & 0.91M       \\
kernel size $\Rightarrow$ 7               & $82.09(0.06)$ & 0.92M       \\
kernel size $\Rightarrow$ 9               & $82.01(0.06)$ & 0.94M       \\
kernel size $\Rightarrow$ 11              & $81.94(0.06)$ & 0.95M       \\
ReLU $\Rightarrow$ GELU                   & $82.39(0.09)$ & 0.95M       \\
fewer activations                         & $83.07(0.12)$ & 0.95M       \\
fewer norms                               & $83.32(0.08)$ & 0.93M       \\
BN $\Rightarrow$ LN                       & $83.39(0.09)$ & 0.95M       \\
GRN module                                & $85.43(0.05)$ & 0.95M       \\
separate d.s. conv(TDResNeXt)             & $85.87(0.07)$ & 1.13M       \\
\hline
\multicolumn{3}{l}{$^{\mathrm{a}}$Reported numbers are $mean(std)$ over five trials}
\end{tabular}
\end{center}
\end{table}

\subsection{Performance Analysis on Training Strategy}
\begin{figure}[htbp]
    \centering
    \begin{subfigure}[b]{0.24\textwidth}
        \centering
        \includegraphics[width=\textwidth]{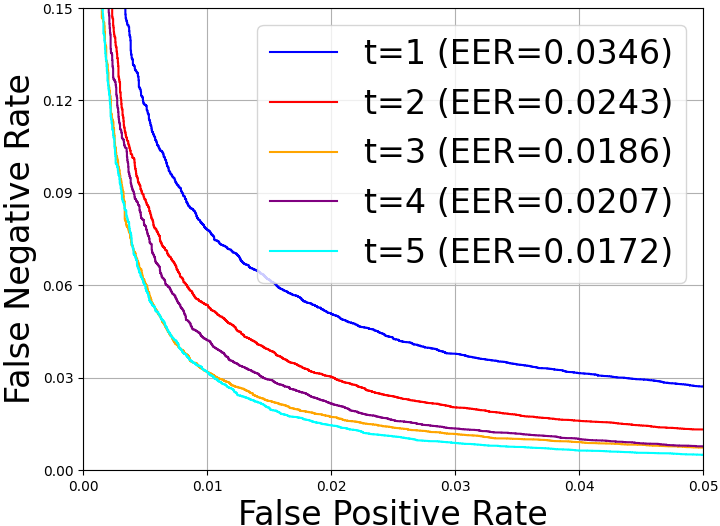}
        \caption{}
        \label{fig:a}
    \end{subfigure}
    \hfill
    \begin{subfigure}[b]{0.24\textwidth}
        \centering
        \includegraphics[width=\textwidth]{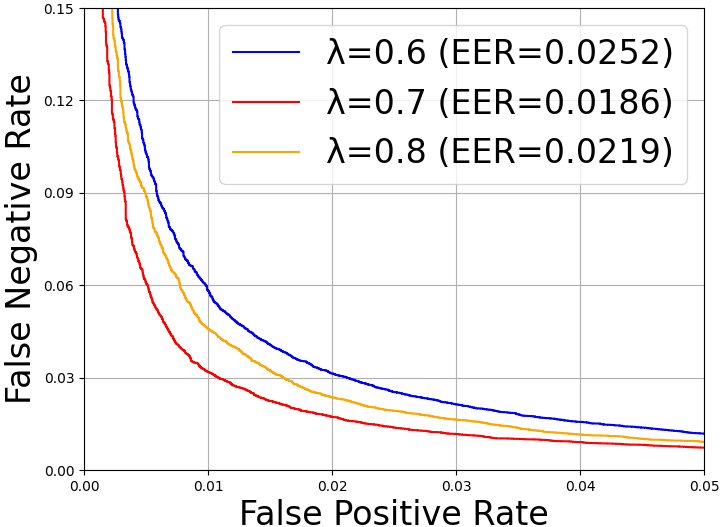}
        \caption{}
        \label{fig:b}
    \end{subfigure}
    \begin{subfigure}[b]{0.24\textwidth}
        \centering
        \includegraphics[width=\textwidth]{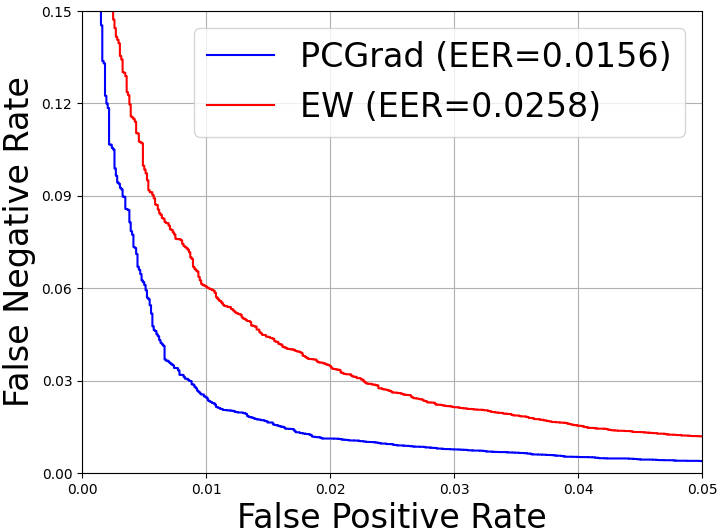}
        \caption{}
        \label{fig:c}
    \end{subfigure}
    \hfill
    \begin{subfigure}[b]{0.24\textwidth}
        \centering
        \includegraphics[width=\textwidth]{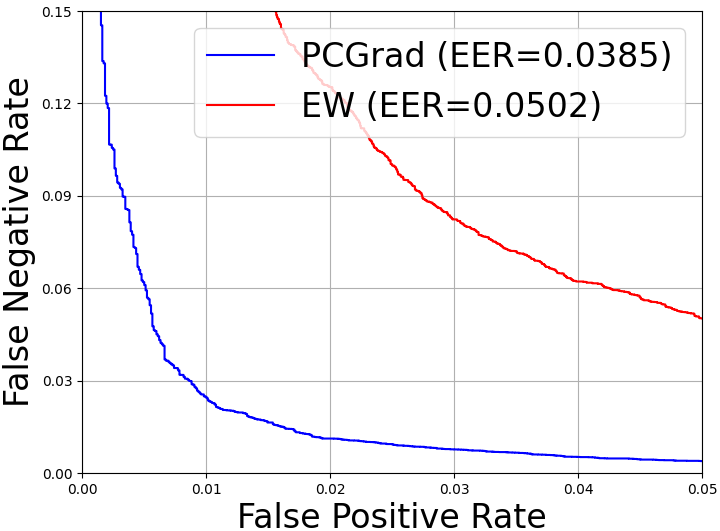}
        \caption{}
        \label{fig:d}
    \end{subfigure}
    \caption{Performance Analysis on Training Strategy}
    \label{fig:combined fnr vs fpr curves}
\end{figure}

As shown in Fig.~\ref{fig:combined fnr vs fpr curves}, we utilize FNR-FPR curves to illustrate the impact of varying parameters and multi-task loss weighting strategies on performance. As discussed previously, the cosine similarity distributions between positive and negative samples must be adjusted by~\eqref{eq:gzt}. 
As depicted in Fig.~\ref{fig:a}, optimal performance is achieved when t = 5, whereas lambda influences the loss weight assigned to positive and negative samples. According to Fig.~\ref{fig:b}, the best performance is achieved when lambda = 0.7. In multi-task learning, the loss gradient directions among tasks may conflict, PCGrad addresses this by projecting the gradients of tasks onto the normal plane of the gradients of other tasks. Fig.~\ref{fig:c} and Fig.~\ref{fig:d} respectively illustrate the performance of PCGrad and EW in the OV-KWS and PCOV-KWS tasks, and PCGrad performs better in both tasks.

\begin{table*}[t]
\caption{Ablation Studies on LibriPhrase Dataset. To evaluate the PCOV-KWS task, we generate LibriPhrase-PCOV ($\textbf{\text{LP}}_\textbf{\text{P}}$) from LibriSpeech~\cite{librispeech}. For the PCOV-KWS task, positive pairs are derived from instances that are both target keywords and target speakers, whereas the remaining samples form negative pairs. }
\begin{center}
\fontsize{9pt}{10pt}\selectfont
\begin{tabular}{|c|c|c|c|c|c|c|c|}
\hline
\multirow{2}{*}{\textbf{Method}} & \multirow{2}{*}{\textbf{Backbone}} & \textbf{SV} & \multicolumn{2}{|c|}{\textbf{OV-KWS}} & \multicolumn{2}{|c|}{\textbf{PCOV-KWS}} & \multirow{2}{*}{\textbf{\#Params}}  \\
\cline{3-7} 
 &  & \textbf{\textit{EER(\%)} $\downarrow$} & \textbf{\textit{AUC(\%)} $\uparrow$} & \textbf{\textit{EER(\%)}} & \textbf{\textit{AUC(\%)}} & \textbf{\textit{EER(\%)}} &  \\
\hline
Vanilla & TC-ResNet14-1.5 & - & 98.94(0.12) & 4.83(0.07) & 51.01(0.12) & 46.52(0.08) & 313k \\
PCOV-KWS & TC-ResNet14-1.5 & 5.89(0.13) & 99.16(0.09) & 4.12(0.06) & 98.32(0.10) & 6.12(0.11) & 326k \\
\hline
Vanilla & TDResNeXt & - & 99.77(0.06) & 1.85(0.06) & 58.63(0.14) & 44.02(0.08) & 198k \\
Vanilla+SV & TDResNeXt & 4.09(0.17) & 99.77(0.06) & 1.85(0.06) & 98.96(0.12) & 4.25(0.13) & 396k \\
PCOV-KWS w/o CIB & TDResNeXt & 4.16(0.09) & 99.85(0.09) & 1.56(0.09) & 99.14(0.09) & 3.98(0.12) & 211k \\
PCOV-KWS & TDResNeXt & \textbf{3.85(0.11)} & \textbf{99.85(0.06)} & \textbf{1.56(0.09)} & \textbf{99.34(0.07)} & \textbf{3.85(0.03)} & 211k \\
\hline
\end{tabular}
\label{tab:ablation study}
\end{center}
\end{table*}

\subsection{Ablation Studies}

\subsubsection{Effectiveness of PCOV-KWS Framework}
As illustrated in the Table~\ref{tab:ablation study}, "Vanilla+SV" indicates that the backbone network is trained separately on KWS and SV tasks before being used to handle different tasks, resulting in increased computational consumption. In comparison to the fifth line, it is evident that the PCOV-KWS architecture demonstrates slight advancements in both SV and OV-KWS tasks, with a notable improvement in the PCOV-KWS task. Furthermore, the second and last rows of the table show that PCOV-KWS enhances all tasks when applied to various backbone networks.

\subsubsection{Effectiveness of TDResNeXt}
Comparing the first and third rows of Table~\ref{tab:ablation study}, it is evident that TDResNeXt demonstrates a significant improvement over TC-ResNet in relation to OV-KWS. Furthermore, comparing the second and fifth rows, a demonstrates superior performance in all tasks when used as an audio encoder in PCOV-KWS, while incurring lower parameter and computational costs.

\subsubsection{Impact of CIB}
From the last two rows, the influence of CIB on PCOV-KWS system is evident. CIB utilizes grid search on the validation set to identify the alpha in~\eqref{eq:CIB} that minimizes EER. PCOV-KWS w/o CIB denotes that we manually set the alpha for various tasks. Specifically, for the PCOV-KWS task, alpha is set to 0.5; for OV-KWS, it is set to 1; and for SV, it is set to 0. It is apparent that the performance declines in both the SV and PCOV-KWS tasks under these configurations.

\subsection{Comparison with Baselines}
\subsubsection{OV-KWS}
We utilize the baselines from~\cite{QbyE-T2-libri, QbyE-T4-phon, CLAD} for comparative analysis. Meanwhile, the evaluation datasets are obtained under identical construction method, enabling an assessment of our proposed method. As shown in Table~\ref{tab:ov-kws}, PCOV-KWS-S was roughly equal to PhonMatchNet on the $\textbf{\text{LP}}_\textbf{\text{H}}$ dataset, whereas PCOV-KWS-M excelled across all tasks.
\subsubsection{C-KWS}
We evaluate the zero-shot capability of our model by comparing the baselines from~\cite{Att_RNN, res15, tcresnet, streaming, MatchBox, TENet, bcresnet, QbyE-T4-phon} with our model on C-KWS task in~\ref{tab:c-kws}. Our model surpassed some full-shot models and demonstrated performance that is close to the SOTA approaches. In addition, PCOV-KWS-M performs better compared to PhonMatchNet, which is also a zero-shot model.
\subsubsection{Analysis on the length of keywords}
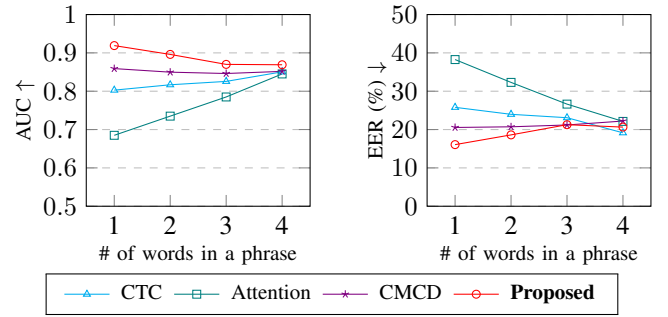
\begin{figure}[h!]
    \begin{minipage}[t]{0.48\columnwidth}
        \centering
        \begin{tikzpicture}
        \pgfplotsset{set layers}
        \begin{axis}[
            scale only axis,
            width=0.7*\linewidth,height=0.6\linewidth,
            xlabel={\# of words in a phrase},
            ylabel={AUC $\uparrow$},
            y label style={at={(0.14,0.5)},font=\footnotesize},
            x label style={at={(0.5,0.05)},font=\footnotesize},
            xmin=0.5, xmax=4.5,
            ymin=0.5, ymax=1.0,
            xtick={1, 2, 3, 4},
            xticklabels={1, 2, 3, 4},
            ytick={0.5, 0.6, 0.7, 0.8, 0.9, 1.0},
            ymajorgrids=true,
            grid style=dashed
        ]
            \addplot[
            color=cyan,
            mark=triangle,
            mark size=1.5
            ]
            coordinates
            {(1,0.8028)(2,0.8168)(3,0.8256)(4,0.8505)};\label{auc_ctc}
            \addplot[
            color=teal,
            mark=square,
            mark size=1.5,
            ]
            coordinates
            {(1,0.685)(2,0.735)(3,0.785)(4,0.845)};\label{auc_att}
            \addplot[
            color=violet,
            mark=star,
            mark size=1.5,
            ]
            coordinates
            {(1,0.859)(2,0.8497)(3,0.8462)(4,0.85175)};\label{auc_CMCD}
            \addplot[
            color=red,
            mark=o,
            mark size=1.5,
            ]
            coordinates
            {(1,0.919)(2,0.896)(3,0.870)(4,0.869)};\label{auc_pro}
        \end{axis}
        \end{tikzpicture}
    \end{minipage}
    \begin{minipage}[t]{0.48\columnwidth}
        \centering
        \begin{tikzpicture}
        \pgfplotsset{set layers}
        \begin{axis}[
            scale only axis,
            width=0.7*\linewidth,height=0.6\linewidth,
            xlabel={\# of words in a phrase},
            ylabel={EER (\%) $\downarrow$},
            y label style={at={(0.19,0.5)},font=\footnotesize},
            x label style={at={(0.5,0.05)},font=\footnotesize},
            xmin=0.5, xmax=4.5,
            ymin=0.0, ymax=50,
            xtick={1, 2, 3, 4},
            xticklabels={1, 2, 3, 4},
            ytick={0, 10, 20, 30, 40, 50},
            ymajorgrids=true,
            grid style=dashed
        ]
            \addplot[
            color=cyan,
            mark=triangle,
            mark size=1.5
            ]
            coordinates
            {(1,25.795)(2,23.975)(3,23.09)(4,19.15)};\label{eer_ctc}
            \addplot[
            color=teal,
            mark=square,
            mark size=1.5,
            ]
            coordinates
            {(1,38.255)(2,32.3)(3,26.635)(4,22.125)};\label{eer_att}
            \addplot[
            color=violet,
            mark=star,
            mark size=1.5,
            ]
            coordinates
            {(1,20.545)(2,20.72)(3,21.195)(4,22.22)};\label{eer_CMCD}
            \addplot[
            color=red,
            mark=o,
            mark size=1.5,
            ]
            coordinates
            {(1,16.07)(2,18.61)(3,21.27)(4,20.64)};\label{eer_pro}
        \end{axis}
        \end{tikzpicture}
    \end{minipage}
    \begin{minipage}[b]{\columnwidth}\fboxsep=0pt\centering
    \fbox{
        \begin{tikzpicture}
        \begin{customlegend}[legend columns=4,legend style={align=left,draw=none,column sep=0.5ex,font=\footnotesize},
                legend entries={CTC,
                                Attention,
                                CMCD,
                                \textbf{Proposed}
                                }]
                \addlegendimage{color=cyan,mark=triangle,mark size=1.5}
                \addlegendimage{color=teal,mark=square,mark size=1.5}
                \addlegendimage{color=violet,mark=star,mark size=1.5}
                \addlegendimage{color=red,mark=o,mark size=1.5}
                \end{customlegend}
        \end{tikzpicture}
    }
    \end{minipage}\hfill
    \vspace{-4pt}
\caption{Evaluation results according to the number of words in a LibriPhrase evaluation set.}
\vspace{-5pt}
\label{fig:length_comparison}
\end{figure}

Fig.~\ref{fig:length_comparison} illustrates how performance varies based on the number of words in a keyword phrase. Our proposed method demonstrates consistently strong detection performance with baselines from~\cite{ctc, QbyE5, QbyE-T2-libri}, regardless of the keyword length.

\begin{table}[t]
\caption{Comparison with Baselines on OV-KWS}
\fontsize{9pt}{10pt}\selectfont
\begin{tabular}{|c|c|cc|cc|}
\hline
\multirow{2}{*}{\textbf{Model}} & \multirow{2}{*}{\textbf{\#Params}} & \multicolumn{2}{c|}{\textbf{\textit{AUC(\%)}} $\uparrow$} & \multicolumn{2}{c|}{\textbf{\textit{EER(\%)}} $\downarrow$} \\
\cline{3-6}
& & $\text{LP}_\text{E}$ & $\text{LP}_\text{H}$ & $\text{LP}_\text{E}$ & $\text{LP}_\text{H}$ \\
\hline
CMCD & 653k & 95.63 & 77.60 & 10.48 & 29.34 \\
CLAD & - & 97.03 & 76.15 & 8.65 & 30.30 \\
PhonMatchNet & 655k & 99.29 & 88.52 & 2.80 & 18.82 \\
PCOV-KWS-S & 211k & 99.77 & 88.50 & 1.85 & 19.33 \\
PCOV-KWS-M & 376k & \textbf{99.88} & \textbf{89.46} & \textbf{1.32} & \textbf{18.38} \\
\hline
\end{tabular}
\label{tab:ov-kws}
\end{table}

\begin{table}[t]
\caption{Comparison with Baselines on C-KWS}
\fontsize{9pt}{10pt}\selectfont
\begin{tabular}{|c|c|c|c|c|}
\hline
\textbf{Model}        & \textbf{0-shot} & \textbf{Acc.(\%)} & \textbf{\#Params} & \textbf{\#FLOPs} \\ \hline
Att-RNN      & \multirow{7}{*}{$\times$} & 95.6     & 202k     & 22.3M   \\ 
ResNet-15    &                        & 95.8     & 238k     & 894M    \\  
TENet12      &                        & 96.6     & 100k     & 2.9M    \\ 
TC-ResNet    &                        & 96.6     & 305k     & 6.7M    \\ 
MHAtt-RNN    &                        & 97.2     & 743k     & 22.7M   \\ 
MatchBoxNet  &                        & 97.5     & 93k      & 11.3M   \\ 
BC-ResNet8   &                        & 98.0     & 312k     & 89.1M   \\ \hline
PhonMatchNet & \multirow{3}{*}{\checkmark}  & 96.8     & 655k     & -       \\ 
PCOV-KWS-S   &                        & 96.6     & 211k     & 1.13M   \\ 
PCOV-KWS-M   &                        & 96.9     & 376k     & 1.8M    \\ \hline
\end{tabular}
\label{tab:c-kws}
\end{table}

\section*{Conclusions}
This study presented a multi-task learning framework for personalized KWS system that leveraged the relationship between keywords and voiceprints of speakers' utterance. We have filled the gap that some multi-task learning frameworks integrating KWS and SV can not detect arbitrary user-defined keywords when distinguishing target users, while maintaining the lightweight and low computational consumption of the network.

\newpage
\clearpage
\bibliographystyle{IEEEtran}
\bibliography{citation}

\end{document}